\documentclass[sn-mathphys,Numbered]{sn-jnl}
\usepackage{graphicx}%
\usepackage{multirow}%
\usepackage{amsmath,amssymb,amsfonts}%
\usepackage{amsthm}%
\usepackage{mathrsfs}%
\usepackage[title]{appendix}%
\usepackage{xcolor}%
\usepackage{textcomp}%
\usepackage{manyfoot}%
\usepackage{booktabs}%
\usepackage{algorithm}%
\usepackage{algorithmicx}%
\usepackage{algpseudocode}%
\usepackage{listings}%
\usepackage[normalem]{ulem}
\newcommand\zfig[1]{{\color{violet}#1}}
\usepackage{times}

\begin{document}

\title[Article Title]
{A Data-Driven Approach to Morphogenesis under Structural Instability}

\author{\fnm{Yingjie} \sur{Zhao}}

\author*{\fnm{Zhiping} \sur{Xu}$^{\ast}$}\email{xuzp@tsinghua.edu.cn}

\affil{\orgdiv{Applied Mechanics Laboratory, Department of Engineering Mechanics}, \orgname{Tsinghua University}, \orgaddress{\city{Beijing} \postcode{100084}, \country{China}}}

\abstract{
Morphological development into evolutionary patterns under structural instability is ubiquitous in living systems and often of vital importance for engineering structures.
Here we propose a data-driven approach to understand and predict their spatiotemporal complexities.
A machine-learning framework is proposed based on the physical modeling of morphogenesis triggered by internal or external forcing.
Digital libraries of structural patterns are constructed from the simulation data, which are then used to recognize the abnormalities, predict their development, and assist in risk assessment and prognosis.
The capabilities to identify the key bifurcation characteristics and predict the history-dependent development from the global and local features are demonstrated by examples of brain growth and aerospace structural design, which offer guidelines for disease diagnosis/prognosis and instability-tolerant design.
}


\maketitle

\clearpage
\newpage

\section*{Introduction}

Morphological evolution and pattern development are key processes of growing living organisms and can be detrimental to engineering structures in service~(\zfig{Fig. 1a})~\cite{chen2023AdvSci,liu2020EML,yin2008PNAS}.
Brain growth from neural tubes into complex cortical folds sees prominent structural evolution.
Cerebral palsy and seizures are highly correlated with abnormal brain morphologies~\cite{severino2020Brain,besson2008Brain,borne2021NeuroImage}.
In engineering, skins of thin-walled aerospace structures buckle under residual compressive stress, resulting in wrinkled morphology and risks of catastrophic structural failure due to abrupt changes in the morphology and load redistribution in the structures~\cite{degenhardt2014design}.
Predicting the loss of stability and subsequent morphological development are of significant importance in disease diagnosis and engineering design.
However, the processes can be complicated by the underlying, often nonlinear processes spanning over multiple lengths and time scales.

The construction of predictive models with simultaneously high accuracy and high efficacy is challenging.
Global features and local characteristic patterns of the brain are inherited from the genetic, molecular, cellular, and mechanical factors in brain development~\cite{klingler2021Science,llinares2019NatureReviewsNeuroscience}.
The nucleation and amplification of wrinkling or buckling patterns in the structural components are relevant to both residual stresses and external loading conditions~\cite{zimmermann2020EFA}.
In practice, domain knowledge is often harnessed to make decisions and predictions based on the development of morphological patterns.
Medical experts diagnose brain abnormalities through visual judgments of neuroimaging features and clinical phenotyping~\cite{severino2020Brain}.
Safety factors are defined as the ratio between the ultimate and limit loads to prevent structural failure in aerospace engineering design ~\cite{bristow2007safety}.

Massive data collected in past decades from experimental tests and high-fidelity computer simulations make it possible to construct databases to assist in disease diagnosis and the design of aerospace structures. 
For brain abnormalities, computed tomography (CT) and magnetic resonance imaging (MRI) provide experimental data of the organ morphologies, which is limited by the resolution of techniques though~\cite{volkow1997PNAS}.
High-fidelity physics-based simulation results contain full-scale information on brain morphology, which are, however, constrained by the experimental facts~\cite{darayi2021review,alenya2022brain_validation,tallinen2016NaturePhysics,wang2021SR}.
The emergence of morphological features such as sulci and gyri indicate large deformation that often leads to geometric, material, and contact nonlinearities in the problem~\cite{hohlfeld2011PRL,holland2018PRL,gandikota2021NJP}.
The transition from smooth to corrugated surfaces is history-dependent, which introduces additional difficulties in the modeling~(\zfig{Supplementary Fig. 1a}).
Recently, machine-learning techniques show the potential to address these issues.
Specially designed network architectures such as the long short-term memory (LSTM) network could tackle the nonlinearities and capture the historical dependency, which is used to learn the process of morphological development and construct the phase diagrams of bifurcation~\cite{hochreiter1997NeuralComputation,hesthaven2018JCP,pichi2021PAMM}.

In aerospace engineering design, thin-walled structures are often used to reduce weight, fuel consumption, and the costs of manufacturing and operating~\cite{degenhardt2014design}.
Plates and cylindrical shells are endangered by buckling withstanding compressive loading~\cite{quinn2009plate,franzoni2019cylinder}.
The morphological development features local, tiny wrinkles after the first buckling, and then intense global crumples that may lead to collapse~\cite{degenhardt2008ComputStruct}.
In conservative design, the ultimate load is commonly set below the first buckling load to prevent instability~\cite{degenhardt2014design}.
Numerically accurate and experimentally validated simulations of post-buckling and structural collapse can elevate the ultimate load beyond the first-buckling load while below the collapse load~\cite{zimmermann2006posicoss,degenhardt2006cocomat}.
Data-driven methods adequately characterize the phase space of structural stability and the bifurcation diagram can forecast the post-buckling and failure behaviors, and allows \emph{instability-tolerant} design reminiscent of the `damage-tolerance' design for structural integrity~(\zfig{Fig. 1b}).
The procedure can be made highly efficient and integrated into the `digital twin' technology provided with high-fidelity databases prepared in advance.

In this work, a machine-learning framework is developed by combining finite element analysis (FEA), the PointNet algorithm, and the LSTM network to resolve the morphological and developmental complexities in the presence of structural instability~(\zfig{Fig. 1c}).
Digital libraries of simulated brain and aerospace structures are constructed using FEA~\cite{tallinen2016NaturePhysics,tallinen2014PNAS,wang2021SR,alenya2022brain_validation,degenhardt2008ComputStruct,degenhardt2014design}.
We use PointNet for feature extraction and dimensionality reduction of the morphological features, and LSTM to capture the historical dependency.
Key indicators such as the gyrification index (GI), mean curvatures, localization factor ($L_{\rm F}$), and dynamic eigenvalues (DEVs) are analyzed.
The information is then fed into tasks including morphological recognition, bifurcation identification, developmental prediction, and coarse-graining of the processes.

\section*{Results}

\subsection*{Digital libraries}
We construct digital libraries of morphological patterns that are then used to identify abnormal patterns, predict their development, and assist in diagnosis and risk reduction~(\zfig{Fig. 1b,c}).
To model brain morphologies, a representative core-shell geometry is used for its simplicity~\cite{tallinen2014PNAS,wang2021SR,xu2022NCS,yin2008PNAS}, as implemented to explore the mechanical instability in cortical folding~\cite{tallinen2016NaturePhysics,striedter2015review,darayi2021review,budday2018IJSS,da2021NeuroImage,alenya2022brain_validation}.
The outer spherical shell represents the cerebral cortex (gray matter), and the inner core for the white matter.
Following the experimental evidences~\cite{fischl2000PNAS,chang2007abnormal,xu2010JBE,dervaux2008PRL}, the cortical thickness ranges from $0.03-1.63$ $\rm{mm}$ according to the abnormal and normal human cerebral cortex measurements and the scale factor~\cite{fischl2000PNAS,chang2007abnormal}, and the relative shear modulus ($\mu_{\rm grey}/\mu_{\rm white}$) ranges from $0.65-1$~\cite{xu2010JBE,dervaux2008PRL}.
The tangential growth (TG) model is used to simulate the cellular mechanisms that create the growth stresses and lead to the pattern evolution (see Methods for details)~\cite{tallinen2014PNAS,tallinen2016NaturePhysics,llinares2019NatureReviewsNeuroscience}.

In aerospace engineering design, thin plates and cylindrical shells are commonly used in primary structures of aircraft and rockets for structural efficiency~\cite{quinn2009plate,franzoni2019cylinder}.
The Johnson-Cook model of plasticity is used for the materials~\cite{johnson1983constitutive,seidt2013plastic}.
Axial compression is applied to trigger buckling of 
the thin-walled structures~\cite{ko2013loadings}.

\subsection*{Morphological recognition}
To resolve the morphological complexity and for diagnosis purposes, the first step is to recognize the geometrical features during brain development, such as the GI and mean curvatures.
We discretize the brain surfaces into a tetrahedral mesh with a density of $10^6$ tetrahedra/cm$^3$~\cite{tallinen2014PNAS,tallinen2016NaturePhysics,wang2021SR}.
The convergence of calculated descriptors is confirmed.
The value of GI measures the depth and width of the cortical folding~\cite{clouchoux2012GI} and is defined as the areal ratio of the cortical surface to the smooth convex hull~(\zfig{Fig. 2a}),
\begin{equation}\label{eq1}
	{\rm GI} = \frac{S_{\rm cortex}}{S_{\rm convex}}.
\end{equation}
The mean curvatures are defined as the average of the two principal curvatures ($k_{\rm max}$ and $k_{\rm min}$)~(\zfig{Supplementary Fig. 2a}),
\begin{equation}\label{eq2}
	\kappa = \frac{k_{\rm max}+k_{\rm min}}{2},
\end{equation}
which quantify the local corrugation of cortical folding.
The average value of absolute curvature, $\lvert \kappa \rvert$, is defined as a key geometrical indicator~\cite{wang2021SR}.
The variance in $\kappa$ increases as the cortical convolution becomes stronger.
In addition to the geometrical measures, topological features are also crucial for neuronal message-passing through chemical and electrical signals across the gyri-to-gyri contact, and vital for brain morphologies and functions~\cite{park2013Science}.
A localization factor ($L_{\rm F}$) is thus defined as the mean distance of contact mapped into the reference configuration to characterize the topology of contact in cortical convolution~\cite{zhao2022Patterns}, that is
\begin{equation}\label{eq3}
	L_{\rm F} = \frac{\sum_{i} \alpha_{i}}{N},
\end{equation}
\begin{equation}\label{eq4}
	\alpha_{i} = \frac{\sum_{j} D_{ij}}{M_{i}},
\end{equation}
where $D_{ij}$ is the distance between the nodes $i$ and $j$ in the mesh grid, $M_{i}$ is the number of contacts for the node $i$, $j$ presents all nodes in contact with $i$, $\alpha_{i}$ represents the average contact distance of the node $i$, and $N$ is the number of all nodes that are in contact with others~(\zfig{Fig. 2b}).

The combined set of parameters (GI, $\kappa$, $L_{\rm F}$) is used to characterize the brain morphology.
Unsupervised clustering of simulated brain morphologies in the space of GI and $\kappa$ is summarized in~\zfig{Fig. 2d}.
Four distinct types are identified and labeled as `convolution-free', `small folds', `thick convolutions', and `normal convolutions', which aligns well with the experimental evidence of lissencephaly, polymicrogyria, pachygyria, and normal brain, respectively~(\zfig{Fig. 2d})~\cite{severino2020Brain}.
The labels are then fed to supervised learning using PointNet.
The confusion matrix of the well-trained classifier which is summarized in \zfig{Fig. 2e,f, Supplementary Fig. 3} shows that the accuracy of morphological identification exceeds $95\%$.

Interestingly, we find that the complexity of pattern development can be coarse-grained according to low-dimensional features such as the network of grooves and ridges in the brain morphology~\cite{boucher2009depth}.
Compared to the original point-cloud representation, the reduced models constructed from the boundaries between grooves and ridges can be used to develop evolutionary dynamics models to understand the underlying biophysical principles.
Local patterns such as the nodes are identified in the groove-and-ridge network~(\zfig{Supplementary Fig. 1b-d}), which are nucleated, grow, and merge, showing distinct bifurcation behaviors.

\subsection*{Pattern development}
To efficiently represent the pattern development, we adopt the PointNet autoencoder by considering rotation and permutation invariance.
The learned representations are then fed to the LSTM network to include the historical dependency in the latent space~(\zfig{Fig. 1c}).
The predicted morphologies at specific stages of development are shown in~(\zfig{Supplementary Fig. 2e,f}) for two representative samples with different cortical thicknesses.
The model predictions are evaluated by the Chamfer distance, which measures the discrepancy between predicted morphologies at different values of observation time $t_{\rm obs}$ and the ground truth~\cite{fan2017ChamferDisance} ~(\zfig{Fig. 2h, Supplementary Fig. 2b, c}).
As $t_{\rm obs}$ increases, more information is extracted and the model corrects earlier predictions, e.g., from cortical convolutions to smooth appearance which forecasts the risk of lissencephaly~(\zfig{Supplementary Fig. 2b,e}).
The models also modify the local patterns continuously, demonstrating the capability to resolve the fine morphological features during the evolution~(\zfig{Fig. 2g,h, Supplementary Fig. 2f}).

Pattern development from smooth to highly curled surfaces involving buckling of the thin shells, which is related to the malformation of the brain~(\zfig{Supplementary Fig. 1b}) resulting from migration and growth of neurons as simulated by FEA using the TG model.
Identifying the bifurcation behaviors in the morphological changes (e.g., the cortical convolutions) is of vital importance for prognosis purposes~(\zfig{Fig. 2i, Supplementary Fig. 2d}).
Our results show that bifurcation behaviors are identified from the evolution of the mean value of $\lvert \kappa \rvert$~(\zfig{Supplementary Fig. 2a}).
Their features are further analyzed through the descriptors such as GI and local $L_{\rm F}$, as well as the DEVs, which characterize the occurrence and nature of bifurcation~(\zfig{Fig. 2a-c, Supplementary Fig. 4,5, see Methods for details})~\cite{grziwotz2023DEV}.
Primary and secondary bifurcation are identified in~\zfig{Supplementary Fig. 5}.
It should be noted that the mapping between the simulation time of brain development and the gestational age is non-linear~\cite{wang2021SR}.
The time series of the global indicator, GI, and DEVs analysis indicate a primary fold bifurcation at $t = 0.725$~(\zfig{Fig. 2a, Supplementary Fig. 5a,b, see Methods for details}).
On the other hand, the time series of the local indicator, $L_{\rm F}$, and DEVs analysis indicate a secondary fold bifurcation at $t = 0.937$~(\zfig{Fig. 2b, Supplementary Fig. 5c,d}).
Impressively, we find that our framework forecasts the bifurcation behaviors at $t = 0.428$ through the Chamfer distance between the prediction and the ground truth~(\zfig{Fig. 2h, Supplementary Fig. 2b,c}).

Moreover, similar to the coarse-grained representation of morphologies in the configurational space, the complexity of developmental patterns in the time domain can be addressed by dimensionality reduction according to the occurrence of bifurcation.
Determining the nature of branching could advance our understanding of the physics that governs the spatiotemporal evolution of complex systems (e.g., for brain development).

\subsection*{Diagnosis of brain malformations}
The well-trained models (recognizer and predictor) based on the digital libraries of the brain can be applied to clinical diagnosis of malformations and abnormal development~(\zfig{Fig. 1c, Supplementary Fig. 3a,b, Supplementary Fig. 6a}).
The point clouds reconstructed from MRI can be fed to the models to assess the current state and predict its future evolution~(\zfig{Fig. 1c, Supplementary Fig. 6a}).
Once transition into a critical branch in the bifurcation diagram is predicted (such as normal brain to lissencephaly, polymicrogyria, or pachygyria), medical intervention may need to be imposed in conjunction with diagnosis by medical experts, forecasting the event and type of bifurcation is thus highly valuable for the diagnosis purposes~(\zfig{Fig. 4a,c}).

\subsection*{Instability-tolerant design for aerospace structures}

Predicting post-buckling behaviors allows to develop instability-tolerant design of aerospace structures~(\zfig{Fig. 3, Supplementary Fig. 6b}).
Once the load increases beyond the buckling limit, plates and cylindrical shells lose their stability through the primary bifurcation, and weak wrinkles form on their surfaces~(\zfig{Fig. 3a}). 
As the load continues to increase, a secondary bifurcation takes place before the collapse.
More significant and irregular crumples form and the structures can no longer withstand the loads~(\zfig{Fig. 3b, Supplementary Fig. 6b}).
In conservative design, limit and ultimate loads are set to be lower than the first buckling load to avoid structural failure~\cite{degenhardt2014design}.
In practice, one can alleviate the design constraint by permitting post-buckling, which reduces the weight and prolong the lifespan of structural components.
Based on accurate and efficient simulations, the European project Improved POstbuckling SImulation for Design of Fibre COmposite Stiffened Fuselage Structures (POSICOSS) lift the limit and ultimate loads above the first buckling load for the fuselage structures, which were validated by comprehensive experimental databases and derived design guidelines~\cite{zimmermann2006posicoss}.
The follow-up project Improved MATerial Exploitation at Safe Design of COmposite Airframe Structures by Accurate Simulation of COllapse (COCOMAT) further accounts for the collapse of structures by considering material degradation~\cite{degenhardt2006cocomat}.
However, these strategies are enforced in the design stage.
An instability-tolerant design proposed in this work can allow for bifurcations with weak effects on the load-bearing capacity~(\zfig{Supplementary Fig. 6b}), and the tolerance of subcritical wrinkles in service reduces the time and cost of maintenance~(\zfig{Fig. 3a}).

\section*{Discussion}

\subsection*{Physics of the model}
In addition to the morphological evolution, internal stress also arises in the structures as driven by mechanical instability.
A recent work studying 2D macromolecular membranes shows that including the lattice strain and topological contact in the machine-learning process significantly improves the performance of morphological classification and identification~\cite{zhao2022Patterns}.
In contrast, the outstanding performance ($>$ $95$\%) reported here using the geometric information only~(\zfig{Fig. 2e and Supplementary Fig. 3c}) suggests that there is an intrinsic relation between morphology and internal stress, while the morphology of 2D macromolecules is sensitive to thermal fluctuation, which is externally perturbed~\cite{wang2020Matter}.

\subsection*{The nature of bifurcations}
The capability to identify bifurcation behaviors helps to construct phase diagrams of structural stability.
Morphological development in complex structures can switch from one branch to another, which is prohibited in traditional path-following methods~(\zfig{Fig. 4c})~\cite{keller1987path-following}.
Fold bifurcation leads to a catastrophic shift, period-doubling results in fluctuation, and Hopf bifurcation drives the system to an oscillatory state~(\zfig{Fig. 4a, Supplementary Fig. 6c}).
Identifying the nature of bifurcation based on dimensionality reduction and LSTM can thus guide the diagnosis and structural health management (SHM) to minimize potential losses~\cite{bury2021PNAS}.
The phase diagram of structural stability explored in the current model is limited by the constructed digital libraries and can be extended by using generative models~\cite{ni2023Chem,luu2023APL}.

\subsection*{Rare external events}
Pattern development under external forcing often experiences uncertain or rare events.
For example, the initial geometry and intrinsic wrinkles may vary from sample to sample.
The cortical thickness may change as a result of neuronal cell migration.
Genes or genomic mutations also modulate the molecular and cellular processes and brain morphology~\cite{barak2011NatureGenetics,poirier2013NatureGenetics}.
For aerospace engineering, bird strikes, conflicts between airplanes, collisions between space debris and satellites, and extreme weather conditions such as lighting strikes or wake vortex circulations also result in uncertain factors that may lead to unexpected morphological pattern evolution and structural failures~\cite{morio2015rare}.
These rare events ~(\zfig{Fig. 4b}) can be handled in the framework by enriching the datasets, using active learning to improve the efficiency of sampling, or reweighting the samples in calculating the loss function~\cite{pickering2022rare_event}.
However, the relatively smooth predictions from data-driven models can underestimate the impact of rare events~\cite{bi2023pangu}.

\section*{Conclusion}
Pattern development due to mechanical instability in living systems and engineering structures usually shows nonlinear dynamics and bifurcation behaviors.
A machine-learning framework is proposed here to resolve progressive morphogenesis.
The performance of our model is demonstrated through two representative problems of brain growth and post-buckling of aerospace thin-walled structures.
The validated approach could assist in clinical diagnosis and prognosis and prevent structural failure in instability-tolerant design based on pre-computed digital libraries.

\clearpage
\newpage

\section*{Methods}

\subsection*{Modeling morphological development}

Brain development is regulated by genetic, molecular, cellular, and mechanical factors across multiple spatiotemporal scales~\cite{klingler2021Science,llinares2019NatureReviewsNeuroscience}, and the differential tangential growth hypothesis is commonly used~\cite{tallinen2016NaturePhysics,klingler2021Science,llinares2019NatureReviewsNeuroscience}.
Finite element analysis (FEA) can model morphological evolution during brain growth at the continuum level~\cite{tallinen2016NaturePhysics,tallinen2014PNAS,darayi2021review,budday2018IJSS,wang2021SR}.
In the TG model, the tangential growth of the outer gray matter is faster than the inner white matter~\cite{tallinen2016NaturePhysics}.
Compression resulting from the mismatch in deformation may then leads to mechanical instability of the brain surface, forming characteristic sulci and gyri structures~\cite{tallinen2014PNAS,tallinen2016NaturePhysics,striedter2015review,darayi2021review,wang2021SR,budday2018IJSS,da2021NeuroImage}.

In continuum modeling, the reference configuration can be mapped to the current one through the deformation gradient tensor as
\begin{equation}\label{eq4}
	\textbf{F} = \textbf{F}^{\rm e}\cdot\textbf{G},
\end{equation}
where $\textbf{F}^{\rm e}$ is the elastic deformation gradient and $\textbf{G}$ is the growth term.
In the TG model, the growth tensor $\textbf{G}$ is
\begin{equation}\label{eq5}
	\textbf{G} = g\textbf{I} + (1-g)\hat{\textbf{n}}\otimes\hat{\textbf{n}},
\end{equation}
where $\hat{\textbf{n}}$ is the surface normal of the reference configuration, $\textbf{I}$ is the unit tensor, and
\begin{equation}\label{eq6}
	g = 1 + \frac{\alpha_{t}}{1+e^{10(\frac{y}{T}-1)}},
\end{equation}
is the growth coefficient, where $\alpha_{t}$ controls the magnitude of local cortical expansion.
There is a smooth transition from the surface of the gray matter layer to the white matter layer with a gradually decreasing growth coefficient.
$y$ is the distance to the surface, and $T$ is the thickness of the cortex.
The brain is modeled as a nonlinear neo-Hookean hyperelastic material, where the strain energy density is
\begin{equation}\label{eq7}
	W = \frac{\mu}{2}[\rm{Tr}(\textbf{F}^{\rm e}\textbf{F}^{\rm eT})\emph{J}^{-2/3}-3] + \frac{\emph{K}}{2}(\emph{J}-1)^2,
\end{equation}
where $\mu$ is the shear modulus, $J$ is the determinant of Jacobian matrix, $K$ is the bulk modulus.
The interaction between surfaces is modeled with an energy penalty via vertex-triangle contact, which prevents the nodes from penetrating the faces of elements~\cite{ericson2004contact}.
The explicit solver is used to minimize the total (elastic and contact) energy of the quasi-static system.
The time step $\Delta t = 0.05a\sqrt{\rho/\emph{K}}$ is set to ensure the convergence, where $a$ is mesh size and $\rho$ is mass density~\cite{belytschko2014FEA}.

\subsection*{Machine learning}

The morphological data obtained from FEA are stored in point clouds, which possess high-dimensional characteristics and redundant information.
Experimentally, point-cloud data can be collected from 3D MRI images~\cite{fan2017MRI}.
Feature extraction and dimensionality reduction are carried out to improve the performance of downstream tasks~(\zfig{Fig. 1c, Supplementary Fig. 7}).
PointNet is used to extract key features and train models for morphological phase classification and recognition~\cite{qi2017pointnet}.
The K-means algorithm is used for clustering, and the labels are fed into PointNet~(\zfig{Fig. 1c}).
The historical dependency of data is addressed by using an autoencoder design based on LSTM~(\zfig{Fig. 1c, Supplementary Fig. 8})~\cite{hochreiter1997NeuralComputation,achlioptas2018ICML}. 

\subsection*{Bifurcation Characterization}

DEVs analysis is performed to characterize the bifurcation behaviors based on Refs.~\cite{grziwotz2023DEV,ushio2018Nature,cenci2019EDM}, which makes use of principal eigenvalues (PEs) of the Jacobian matrix in the reconstructed state space of a univariate time series~\cite{grziwotz2023DEV}.
A dynamical system approaches the bifurcation point as the Euclidean norm of PEs approaches $1$ (\zfig{Fig. 4a, Supplementary Fig. 6c})~\cite{grziwotz2023DEV,kuznetsov1998bifurcation}.
According to the characteristics of PEs in the complex plane, the type of bifurcation can be determined (\zfig{Fig. 4a, Supplementary Fig. 6c})~\cite{grziwotz2023DEV,kuznetsov1998bifurcation}.
Specifically, a fold bifurcation has a real part of $1$ and an imaginary part of $0$.
The period-doubling bifurcation can be recognized by real and imaginary parts of $-1$ and $0$, respectively.
The Hopf bifurcation has a non-zero imaginary part and the Euclidean norm of PEs approaches $1$~\cite{grziwotz2023DEV,kuznetsov1998bifurcation}.

\clearpage
\newpage

\begin{figure}[htb]
\centering
\includegraphics[width=13cm] {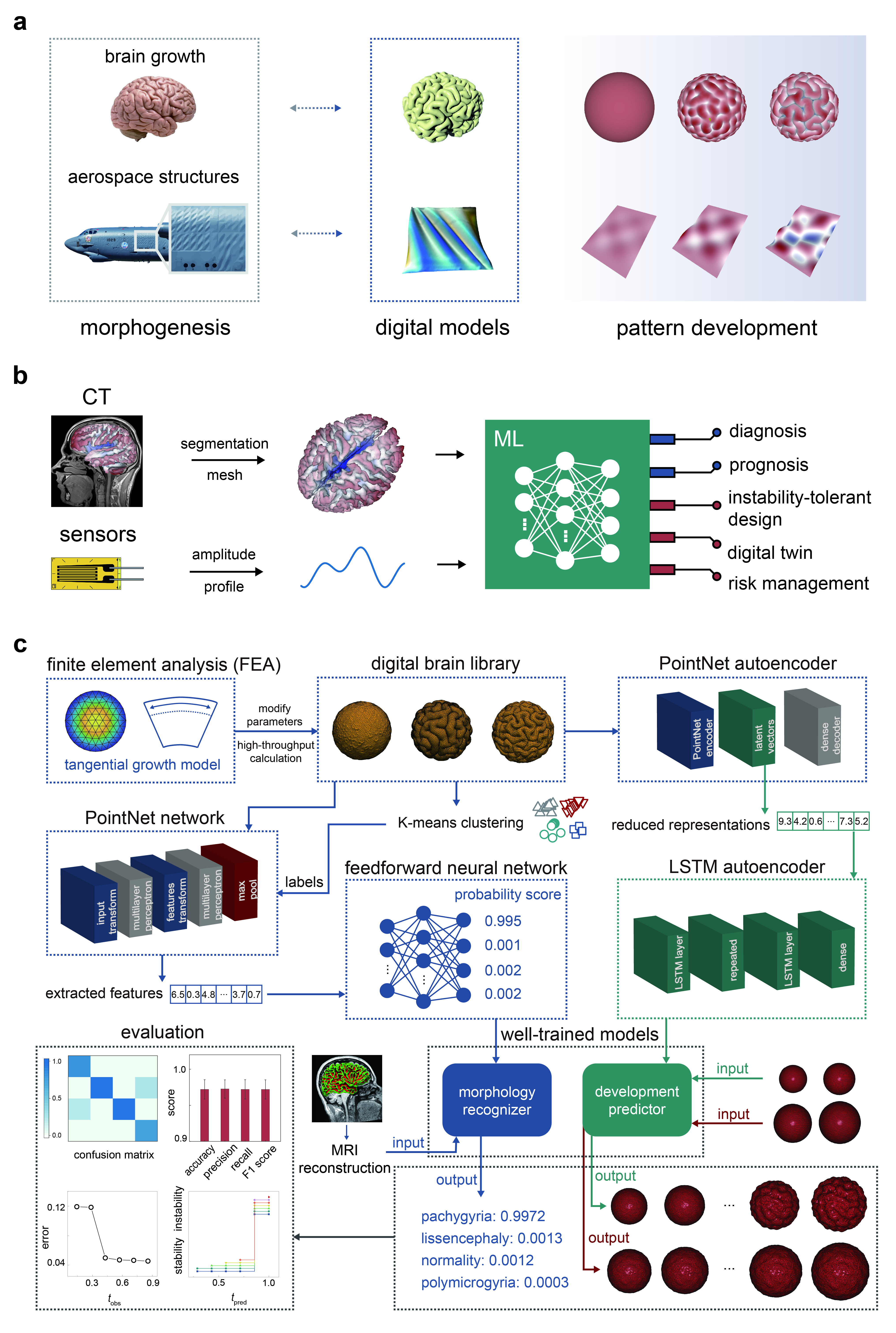} 
\caption{ {\bf The machine-learning framework for pattern development.}
{\bf a}, Morphogenesis, digital models, and pattern development.
{\bf b}, Disease diagnosis and instability-tolerant design in the brain growth and aerospace
structural design.
{\bf c}, The neural-network framework to address the spatiotemporal complexities of morphogenesis.
}
\label {fig1}
\end{figure}

\clearpage
\newpage

\begin{figure}[htb]
\centering
\includegraphics[width=13cm] {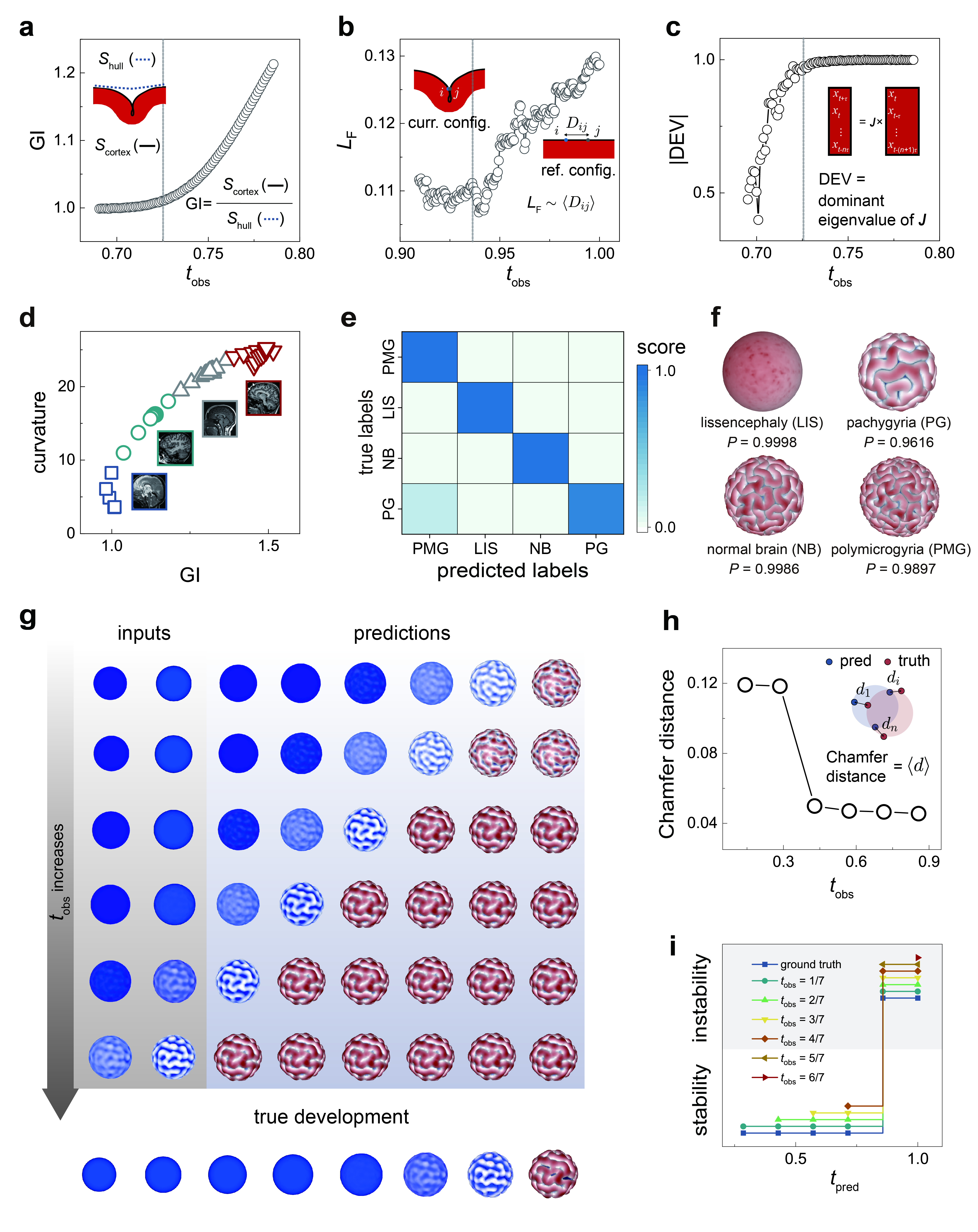} 
\caption{ {\bf Identification and prediction of morphological development.}
{\bf a-c}, Sequences of structural features involving global feature gyrification index (GI) ({\bf a}), local feature localization factor ($L_{\rm F}$) ({\bf b}), and the absolute values of dynamical eigenvalues (DEVs) ({\bf c}).
{\bf d}, Unsupervised clustering of morphologies in the space spanned by GI and curvature.
{\bf e}, The confusion matrix of recognized brain morphologies.
{\bf f}, The representative morphological samples and corresponding probabilities predicted by the models, in which a high probability value indicates the ability of models to identify the morphology of the simulated brain.
{\bf g}, Predicted map of brain growth.
{\bf h}, The Chamfer distance calculated between predicted final morphology and the ground truth as $t_{\rm obs}$ increases.
{\bf i}, The branching behaviors can be accurately characterized during brain growth.
}
\label {fig2}
\end{figure}

\clearpage
\newpage

\begin{figure}[htb]
\centering
\includegraphics[width=13cm] {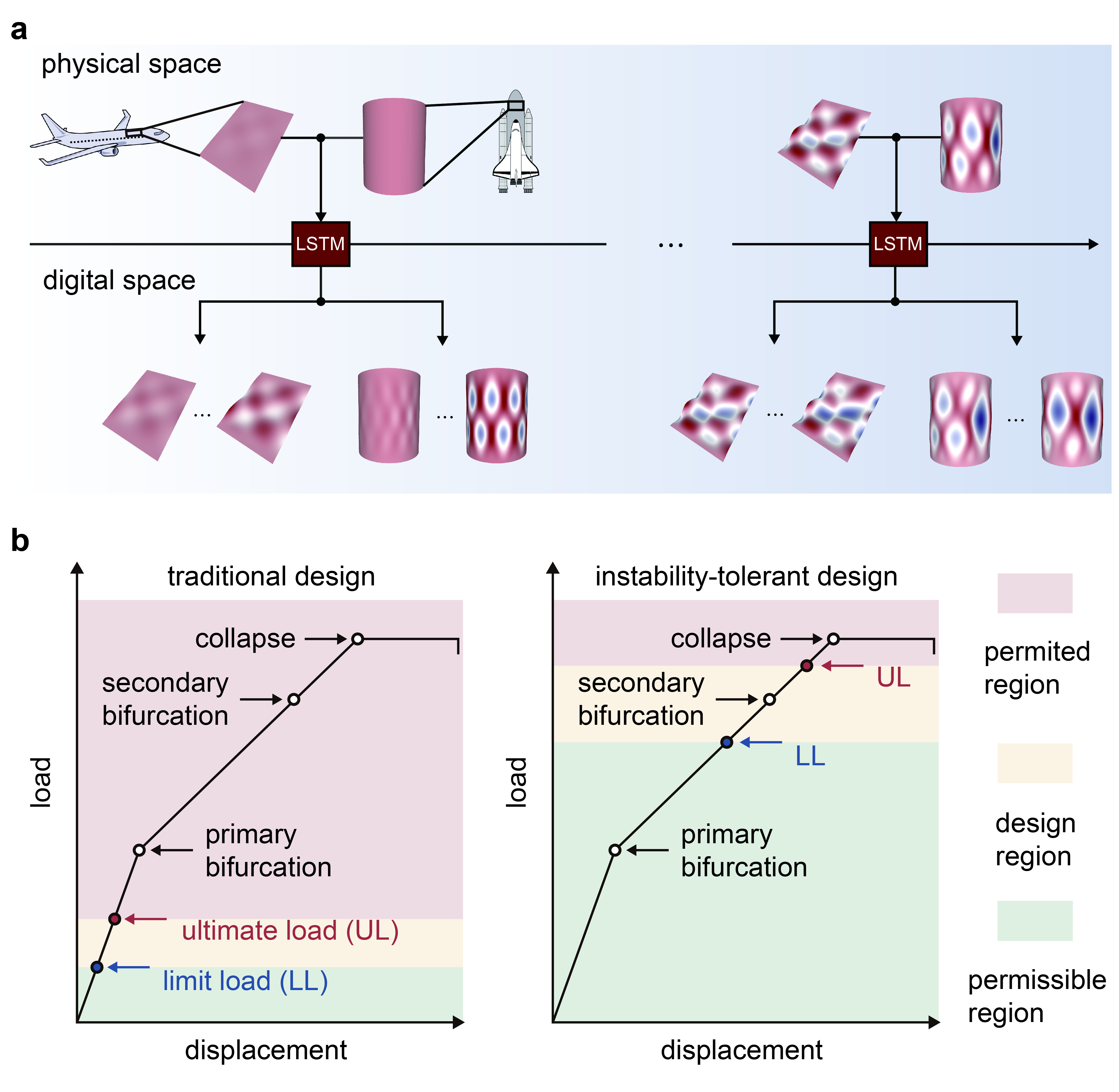} 
\caption{ {\bf Instability-tolerant design for aerospace structures.}
{\bf a}, Real-time predictions of pattern evolution for aerospace structures.
{\bf b}, Instability-tolerant design enlarges the permissible regions in design in comparison with traditional design.
}
\label {fig3}
\end{figure}

\clearpage
\newpage

\begin{figure}[htb]
\centering
\includegraphics[width=13cm] {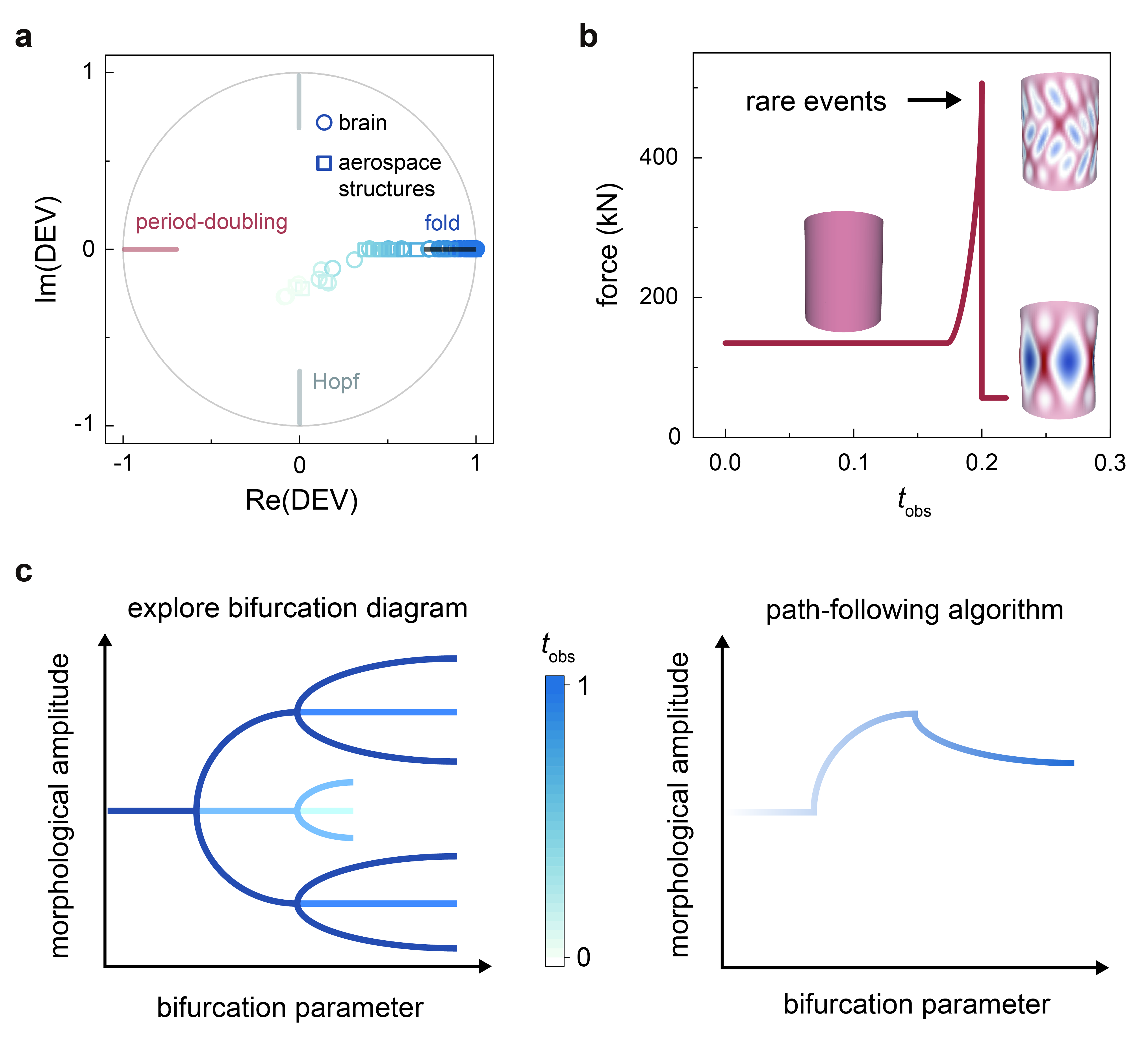} 
\caption{ {\bf Bifurcation behaviors and rare events.}
{\bf a}, The bifurcation nature of pattern development for brain growth and aerospace structural design is identified as fold bifurcation by the analysis of DEVs.
{\bf b}, Rare events generate complex morphological patterns and structural failures.
{\bf c}, The bifurcation diagram covered by the constructed digital libraries can be fully explored, in stark contrast with the single-path nature in path-following methods~\cite{keller1987path-following}.
}
\label {fig4}
\end{figure}

\clearpage
\newpage

\section*{Data availability}

All data needed to evaluate the conclusions are present in the paper and/or the Supplementary Materials.
Additional data related to this paper may be requested from the authors.

\section*{Code availability}
The code used for this study is available at \href{https://github.com/zhaoyj21/instability_tolerance}{https://github.com/zhaoyj21/instability\_tolerance} and the Zenodo repository 
\href{https://zenodo.org/badge/latestdoi/657911680}{(https://zenodo.org/badge/latestdoi/657911680)}.


\bibliography{main_text}


\begin{thebibliography}{63}
\ifx \bisbn   \undefined \def \bisbn  #1{ISBN #1}\fi
\ifx \binits  \undefined \def \binits#1{#1}\fi
\ifx \bauthor  \undefined \def \bauthor#1{#1}\fi
\ifx \batitle  \undefined \def \batitle#1{#1}\fi
\ifx \bjtitle  \undefined \def \bjtitle#1{#1}\fi
\ifx \bvolume  \undefined \def \bvolume#1{\textbf{#1}}\fi
\ifx \byear  \undefined \def \byear#1{#1}\fi
\ifx \bissue  \undefined \def \bissue#1{#1}\fi
\ifx \bfpage  \undefined \def \bfpage#1{#1}\fi
\ifx \blpage  \undefined \def \blpage #1{#1}\fi
\ifx \burl  \undefined \def \burl#1{\textsf{#1}}\fi
\ifx \doiurl  \undefined \def \doiurl#1{\url{https://doi.org/#1}}\fi
\ifx \betal  \undefined \def \betal{\textit{et al.}}\fi
\ifx \binstitute  \undefined \def \binstitute#1{#1}\fi
\ifx \binstitutionaled  \undefined \def \binstitutionaled#1{#1}\fi
\ifx \bctitle  \undefined \def \bctitle#1{#1}\fi
\ifx \beditor  \undefined \def \beditor#1{#1}\fi
\ifx \bpublisher  \undefined \def \bpublisher#1{#1}\fi
\ifx \bbtitle  \undefined \def \bbtitle#1{#1}\fi
\ifx \bedition  \undefined \def \bedition#1{#1}\fi
\ifx \bseriesno  \undefined \def \bseriesno#1{#1}\fi
\ifx \blocation  \undefined \def \blocation#1{#1}\fi
\ifx \bsertitle  \undefined \def \bsertitle#1{#1}\fi
\ifx \bsnm \undefined \def \bsnm#1{#1}\fi
\ifx \bsuffix \undefined \def \bsuffix#1{#1}\fi
\ifx \bparticle \undefined \def \bparticle#1{#1}\fi
\ifx \barticle \undefined \def \barticle#1{#1}\fi
\bibcommenthead
\ifx \bconfdate \undefined \def \bconfdate #1{#1}\fi
\ifx \botherref \undefined \def \botherref #1{#1}\fi
\ifx \url \undefined \def \url#1{\textsf{#1}}\fi
\ifx \bchapter \undefined \def \bchapter#1{#1}\fi
\ifx \bbook \undefined \def \bbook#1{#1}\fi
\ifx \bcomment \undefined \def \bcomment#1{#1}\fi
\ifx \oauthor \undefined \def \oauthor#1{#1}\fi
\ifx \citeauthoryear \undefined \def \citeauthoryear#1{#1}\fi
\ifx \endbibitem  \undefined \def \endbibitem {}\fi
\ifx \bconflocation  \undefined \def \bconflocation#1{#1}\fi
\ifx \arxivurl  \undefined \def \arxivurl#1{\textsf{#1}}\fi
\csname PreBibitemsHook\endcsname

\bibitem[\protect\citeauthoryear{Chen and Gu}{2023}]{chen2023AdvSci}
\begin{barticle}
\bauthor{\bsnm{Chen}, \binits{C.-T.}},
\bauthor{\bsnm{Gu}, \binits{G.X.}}:
\batitle{Physics-informed deep-learning for elasticity: Forward, inverse, and
  mixed problems}.
\bjtitle{Adv. Sci.}
\bvolume{10},
\bfpage{2300439}
(\byear{2023})
\end{barticle}
\endbibitem

\bibitem[\protect\citeauthoryear{Liu et~al.}{2020}]{liu2020EML}
\begin{barticle}
\bauthor{\bsnm{Liu}, \binits{F.}},
\bauthor{\bsnm{Jiang}, \binits{X.}},
\bauthor{\bsnm{Wang}, \binits{X.}},
\bauthor{\bsnm{Wang}, \binits{L.}}:
\batitle{Machine learning-based design and optimization of curved beams for
  multistable structures and metamaterials}.
\bjtitle{Extreme Mech. Lett.}
\bvolume{41},
\bfpage{101002}
(\byear{2020})
\end{barticle}
\endbibitem

\bibitem[\protect\citeauthoryear{Yin et~al.}{2008}]{yin2008PNAS}
\begin{barticle}
\bauthor{\bsnm{Yin}, \binits{J.}},
\bauthor{\bsnm{Cao}, \binits{Z.}},
\bauthor{\bsnm{Li}, \binits{C.}},
\bauthor{\bsnm{Sheinman}, \binits{I.}},
\bauthor{\bsnm{Chen}, \binits{X.}}:
\batitle{Stress-driven buckling patterns in spheroidal core/shell structures}.
\bjtitle{Proc. Natl. Acad. Sci. USA}
\bvolume{105}(\bissue{49}),
\bfpage{19132}--\blpage{19135}
(\byear{2008})
\end{barticle}
\endbibitem

\bibitem[\protect\citeauthoryear{Severino et~al.}{2020}]{severino2020Brain}
\begin{barticle}
\bauthor{\bsnm{Severino}, \binits{M.}},
\bauthor{\bsnm{Geraldo}, \binits{A.F.}},
\bauthor{\bsnm{Utz}, \binits{N.}},
\bauthor{\bsnm{Tortora}, \binits{D.}},
\bauthor{\bsnm{Pogledic}, \binits{I.}},
\bauthor{\bsnm{Klonowski}, \binits{W.}},
\bauthor{\bsnm{Triulzi}, \binits{F.}},
\bauthor{\bsnm{Arrigoni}, \binits{F.}},
\bauthor{\bsnm{Mankad}, \binits{K.}},
\bauthor{\bsnm{Leventer}, \binits{R.J.}}, \betal:
\batitle{Definitions and classification of malformations of cortical
  development: practical guidelines}.
\bjtitle{Brain}
\bvolume{143}(\bissue{10}),
\bfpage{2874}--\blpage{2894}
(\byear{2020})
\end{barticle}
\endbibitem

\bibitem[\protect\citeauthoryear{Besson et~al.}{2008}]{besson2008Brain}
\begin{barticle}
\bauthor{\bsnm{Besson}, \binits{P.}},
\bauthor{\bsnm{Andermann}, \binits{F.}},
\bauthor{\bsnm{Dubeau}, \binits{F.}},
\bauthor{\bsnm{Bernasconi}, \binits{A.}}:
\batitle{Small focal cortical dysplasia lesions are located at the bottom of a
  deep sulcus}.
\bjtitle{Brain}
\bvolume{131}(\bissue{12}),
\bfpage{3246}--\blpage{3255}
(\byear{2008})
\end{barticle}
\endbibitem

\bibitem[\protect\citeauthoryear{Borne et~al.}{2021}]{borne2021NeuroImage}
\begin{barticle}
\bauthor{\bsnm{Borne}, \binits{L.}},
\bauthor{\bsnm{Rivi{\`e}re}, \binits{D.}},
\bauthor{\bsnm{Cachia}, \binits{A.}},
\bauthor{\bsnm{Roca}, \binits{P.}},
\bauthor{\bsnm{Mellerio}, \binits{C.}},
\bauthor{\bsnm{Oppenheim}, \binits{C.}},
\bauthor{\bsnm{Mangin}, \binits{J.-F.}}:
\batitle{Automatic recognition of specific local cortical folding patterns}.
\bjtitle{NeuroImage}
\bvolume{238},
\bfpage{118208}
(\byear{2021})
\end{barticle}
\endbibitem

\bibitem[\protect\citeauthoryear{Degenhardt
  et~al.}{2014}]{degenhardt2014design}
\begin{barticle}
\bauthor{\bsnm{Degenhardt}, \binits{R.}},
\bauthor{\bsnm{Castro}, \binits{S.G.}},
\bauthor{\bsnm{Arbelo}, \binits{M.A.}},
\bauthor{\bsnm{Zimmerman}, \binits{R.}},
\bauthor{\bsnm{Khakimova}, \binits{R.}},
\bauthor{\bsnm{Kling}, \binits{A.}}:
\batitle{Future structural stability design for composite space and airframe
  structures}.
\bjtitle{Thin-Walled Struct.}
\bvolume{81},
\bfpage{29}--\blpage{38}
(\byear{2014})
\end{barticle}
\endbibitem

\bibitem[\protect\citeauthoryear{Klingler et~al.}{2021}]{klingler2021Science}
\begin{barticle}
\bauthor{\bsnm{Klingler}, \binits{E.}},
\bauthor{\bsnm{Francis}, \binits{F.}},
\bauthor{\bsnm{Jabaudon}, \binits{D.}},
\bauthor{\bsnm{Cappello}, \binits{S.}}:
\batitle{Mapping the molecular and cellular complexity of cortical
  malformations}.
\bjtitle{Science}
\bvolume{371}(\bissue{6527}),
\bfpage{4517}
(\byear{2021})
\end{barticle}
\endbibitem

\bibitem[\protect\citeauthoryear{Llinares-Benadero and
  Borrell}{2019}]{llinares2019NatureReviewsNeuroscience}
\begin{barticle}
\bauthor{\bsnm{Llinares-Benadero}, \binits{C.}},
\bauthor{\bsnm{Borrell}, \binits{V.}}:
\batitle{Deconstructing cortical folding: {G}enetic, cellular and mechanical
  determinants}.
\bjtitle{Nat. Rev. Neurosci.}
\bvolume{20}(\bissue{3}),
\bfpage{161}--\blpage{176}
(\byear{2019})
\end{barticle}
\endbibitem

\bibitem[\protect\citeauthoryear{Zimmermann and Wang}{2020}]{zimmermann2020EFA}
\begin{barticle}
\bauthor{\bsnm{Zimmermann}, \binits{N.}},
\bauthor{\bsnm{Wang}, \binits{P.H.}}:
\batitle{A review of failure modes and fracture analysis of aircraft composite
  materials}.
\bjtitle{Eng. Fail. Anal.}
\bvolume{115},
\bfpage{104692}
(\byear{2020})
\end{barticle}
\endbibitem

\bibitem[\protect\citeauthoryear{Bristow and Irving}{2007}]{bristow2007safety}
\begin{barticle}
\bauthor{\bsnm{Bristow}, \binits{J.W.}},
\bauthor{\bsnm{Irving}, \binits{P.}}:
\batitle{Safety factors in civil aircraft design requirements}.
\bjtitle{Eng. Fail. Anal.}
\bvolume{14}(\bissue{3}),
\bfpage{459}--\blpage{470}
(\byear{2007})
\end{barticle}
\endbibitem

\bibitem[\protect\citeauthoryear{Volkow et~al.}{1997}]{volkow1997PNAS}
\begin{barticle}
\bauthor{\bsnm{Volkow}, \binits{N.D.}},
\bauthor{\bsnm{Rosen}, \binits{B.}},
\bauthor{\bsnm{Farde}, \binits{L.}}:
\batitle{Imaging the living human brain: {M}agnetic resonance imaging and
  positron emission tomography}.
\bjtitle{Proc. Natl. Acad. Sci. USA}
\bvolume{94}(\bissue{7}),
\bfpage{2787}--\blpage{2788}
(\byear{1997})
\end{barticle}
\endbibitem

\bibitem[\protect\citeauthoryear{Darayi et~al.}{2021}]{darayi2021review}
\begin{botherref}
\oauthor{\bsnm{Darayi}, \binits{M.}},
\oauthor{\bsnm{Hoffman}, \binits{M.E.}},
\oauthor{\bsnm{Sayut}, \binits{J.}},
\oauthor{\bsnm{Wang}, \binits{S.}},
\oauthor{\bsnm{Demirci}, \binits{N.}},
\oauthor{\bsnm{Consolini}, \binits{J.}},
\oauthor{\bsnm{Holland}, \binits{M.A.}}:
Computational models of cortical folding: A review of common approaches.
J. Biomech.,
110851
(2021)
\end{botherref}
\endbibitem

\bibitem[\protect\citeauthoryear{Aleny{\`a}
  et~al.}{2022}]{alenya2022brain_validation}
\begin{barticle}
\bauthor{\bsnm{Aleny{\`a}}, \binits{M.}},
\bauthor{\bsnm{Wang}, \binits{X.}},
\bauthor{\bsnm{Lef{\`e}vre}, \binits{J.}},
\bauthor{\bsnm{Auzias}, \binits{G.}},
\bauthor{\bsnm{Fouquet}, \binits{B.}},
\bauthor{\bsnm{Eixarch}, \binits{E.}},
\bauthor{\bsnm{Rousseau}, \binits{F.}},
\bauthor{\bsnm{Camara}, \binits{O.}}:
\batitle{Computational pipeline for the generation and validation of
  patient-specific mechanical models of brain development}.
\bjtitle{Brain Multiphys.}
\bvolume{3},
\bfpage{100045}
(\byear{2022})
\end{barticle}
\endbibitem

\bibitem[\protect\citeauthoryear{Tallinen
  et~al.}{2016}]{tallinen2016NaturePhysics}
\begin{barticle}
\bauthor{\bsnm{Tallinen}, \binits{T.}},
\bauthor{\bsnm{Chung}, \binits{J.Y.}},
\bauthor{\bsnm{Rousseau}, \binits{F.}},
\bauthor{\bsnm{Girard}, \binits{N.}},
\bauthor{\bsnm{Lef{\`e}vre}, \binits{J.}},
\bauthor{\bsnm{Mahadevan}, \binits{L.}}:
\batitle{On the growth and form of cortical convolutions}.
\bjtitle{Nat. Phys.}
\bvolume{12}(\bissue{6}),
\bfpage{588}--\blpage{593}
(\byear{2016})
\end{barticle}
\endbibitem

\bibitem[\protect\citeauthoryear{Wang et~al.}{2021}]{wang2021SR}
\begin{barticle}
\bauthor{\bsnm{Wang}, \binits{X.}},
\bauthor{\bsnm{Lef{\`e}vre}, \binits{J.}},
\bauthor{\bsnm{Bohi}, \binits{A.}},
\bauthor{\bsnm{Harrach}, \binits{M.A.}},
\bauthor{\bsnm{Dinomais}, \binits{M.}},
\bauthor{\bsnm{Rousseau}, \binits{F.}}:
\batitle{The influence of biophysical parameters in a biomechanical model of
  cortical folding patterns}.
\bjtitle{Sci. Rep.}
\bvolume{11},
\bfpage{7686}
(\byear{2021})
\end{barticle}
\endbibitem

\bibitem[\protect\citeauthoryear{Hohlfeld and
  Mahadevan}{2011}]{hohlfeld2011PRL}
\begin{barticle}
\bauthor{\bsnm{Hohlfeld}, \binits{E.}},
\bauthor{\bsnm{Mahadevan}, \binits{L.}}:
\batitle{Unfolding the sulcus}.
\bjtitle{Phys. Rev. Lett.}
\bvolume{106}(\bissue{10}),
\bfpage{105702}
(\byear{2011})
\end{barticle}
\endbibitem

\bibitem[\protect\citeauthoryear{Holland et~al.}{2018}]{holland2018PRL}
\begin{barticle}
\bauthor{\bsnm{Holland}, \binits{M.}},
\bauthor{\bsnm{Budday}, \binits{S.}},
\bauthor{\bsnm{Goriely}, \binits{A.}},
\bauthor{\bsnm{Kuhl}, \binits{E.}}:
\batitle{Symmetry breaking in wrinkling patterns: Gyri are universally thicker
  than sulci}.
\bjtitle{Phys. Rev. Lett.}
\bvolume{121}(\bissue{22}),
\bfpage{228002}
(\byear{2018})
\end{barticle}
\endbibitem

\bibitem[\protect\citeauthoryear{Gandikota and
  Schwarz}{2021}]{gandikota2021NJP}
\begin{barticle}
\bauthor{\bsnm{Gandikota}, \binits{M.}},
\bauthor{\bsnm{Schwarz}, \binits{J.}}:
\batitle{Buckling without bending morphogenesis: {N}onlinearities, spatial
  confinement, and branching hierarchies}.
\bjtitle{New J. Phys.}
\bvolume{23}(\bissue{6}),
\bfpage{063060}
(\byear{2021})
\end{barticle}
\endbibitem

\bibitem[\protect\citeauthoryear{Hochreiter and
  Schmidhuber}{1997}]{hochreiter1997NeuralComputation}
\begin{barticle}
\bauthor{\bsnm{Hochreiter}, \binits{S.}},
\bauthor{\bsnm{Schmidhuber}, \binits{J.}}:
\batitle{Long short-term memory}.
\bjtitle{Neural Comput.}
\bvolume{9}(\bissue{8}),
\bfpage{1735}--\blpage{1780}
(\byear{1997})
\end{barticle}
\endbibitem

\bibitem[\protect\citeauthoryear{Hesthaven and
  Ubbiali}{2018}]{hesthaven2018JCP}
\begin{barticle}
\bauthor{\bsnm{Hesthaven}, \binits{J.S.}},
\bauthor{\bsnm{Ubbiali}, \binits{S.}}:
\batitle{Non-intrusive reduced order modeling of nonlinear problems using
  neural networks}.
\bjtitle{J. Comput. Phys.}
\bvolume{363},
\bfpage{55}--\blpage{78}
(\byear{2018})
\end{barticle}
\endbibitem

\bibitem[\protect\citeauthoryear{Pichi et~al.}{2021}]{pichi2021PAMM}
\begin{barticle}
\bauthor{\bsnm{Pichi}, \binits{F.}},
\bauthor{\bsnm{Ballarin}, \binits{F.}},
\bauthor{\bsnm{Rozza}, \binits{G.}},
\bauthor{\bsnm{Hesthaven}, \binits{J.S.}}:
\batitle{Artificial neural network for bifurcating phenomena modelled by
  nonlinear parametrized pdes}.
\bjtitle{PAMM}
\bvolume{20}(\bissue{S1}),
\bfpage{202000350}
(\byear{2021})
\end{barticle}
\endbibitem

\bibitem[\protect\citeauthoryear{Quinn et~al.}{2009}]{quinn2009plate}
\begin{barticle}
\bauthor{\bsnm{Quinn}, \binits{D.}},
\bauthor{\bsnm{Murphy}, \binits{A.}},
\bauthor{\bsnm{McEwan}, \binits{W.}},
\bauthor{\bsnm{Lemaitre}, \binits{F.}}:
\batitle{Stiffened panel stability behaviour and performance gains with plate
  prismatic sub-stiffening}.
\bjtitle{Thin-Walled Struct.}
\bvolume{47}(\bissue{12}),
\bfpage{1457}--\blpage{1468}
(\byear{2009})
\end{barticle}
\endbibitem

\bibitem[\protect\citeauthoryear{Franzoni et~al.}{2019}]{franzoni2019cylinder}
\begin{barticle}
\bauthor{\bsnm{Franzoni}, \binits{F.}},
\bauthor{\bsnm{Odermann}, \binits{F.}},
\bauthor{\bsnm{Wilckens}, \binits{D.}},
\bauthor{\bsnm{Sku{\c{k}}is}, \binits{E.}},
\bauthor{\bsnm{Kalni{\c{n}}{\v{s}}}, \binits{K.}},
\bauthor{\bsnm{Arbelo}, \binits{M.A.}},
\bauthor{\bsnm{Degenhardt}, \binits{R.}}:
\batitle{Assessing the axial buckling load of a pressurized orthotropic
  cylindrical shell through vibration correlation technique}.
\bjtitle{Thin-Walled Struct.}
\bvolume{137},
\bfpage{353}--\blpage{366}
(\byear{2019})
\end{barticle}
\endbibitem

\bibitem[\protect\citeauthoryear{Degenhardt
  et~al.}{2008}]{degenhardt2008ComputStruct}
\begin{barticle}
\bauthor{\bsnm{Degenhardt}, \binits{R.}},
\bauthor{\bsnm{Kling}, \binits{A.}},
\bauthor{\bsnm{Rohwer}, \binits{K.}},
\bauthor{\bsnm{Orifici}, \binits{A.}},
\bauthor{\bsnm{Thomson}, \binits{R.}}:
\batitle{Design and analysis of stiffened composite panels including
  post-buckling and collapse}.
\bjtitle{Comput. Struct.}
\bvolume{86}(\bissue{9}),
\bfpage{919}--\blpage{929}
(\byear{2008})
\end{barticle}
\endbibitem

\bibitem[\protect\citeauthoryear{Zimmermann and
  Rolfes}{2006}]{zimmermann2006posicoss}
\begin{barticle}
\bauthor{\bsnm{Zimmermann}, \binits{R.}},
\bauthor{\bsnm{Rolfes}, \binits{R.}}:
\batitle{P{OSICOSS}: {I}mproved postbuckling simulation for design of fibre
  composite stiffened fuselage structures}.
\bjtitle{Compos. Struct.}
\bvolume{73}(\bissue{2}),
\bfpage{171}--\blpage{174}
(\byear{2006})
\end{barticle}
\endbibitem

\bibitem[\protect\citeauthoryear{Degenhardt
  et~al.}{2006}]{degenhardt2006cocomat}
\begin{barticle}
\bauthor{\bsnm{Degenhardt}, \binits{R.}},
\bauthor{\bsnm{Rolfes}, \binits{R.}},
\bauthor{\bsnm{Zimmermann}, \binits{R.}},
\bauthor{\bsnm{Rohwer}, \binits{K.}}:
\batitle{C{OCOMAT}: {I}mproved material exploitation of composite airframe
  structures by accurate simulation of postbuckling and collapse}.
\bjtitle{Compos. Struct.}
\bvolume{73}(\bissue{2}),
\bfpage{175}--\blpage{178}
(\byear{2006})
\end{barticle}
\endbibitem

\bibitem[\protect\citeauthoryear{Tallinen et~al.}{2014}]{tallinen2014PNAS}
\begin{barticle}
\bauthor{\bsnm{Tallinen}, \binits{T.}},
\bauthor{\bsnm{Chung}, \binits{J.Y.}},
\bauthor{\bsnm{Biggins}, \binits{J.S.}},
\bauthor{\bsnm{Mahadevan}, \binits{L.}}:
\batitle{Gyrification from constrained cortical expansion}.
\bjtitle{Proc. Natl. Acad. Sci. USA}
\bvolume{111}(\bissue{35}),
\bfpage{12667}--\blpage{12672}
(\byear{2014})
\end{barticle}
\endbibitem

\bibitem[\protect\citeauthoryear{Xu et~al.}{2022}]{xu2022NCS}
\begin{barticle}
\bauthor{\bsnm{Xu}, \binits{F.}},
\bauthor{\bsnm{Huang}, \binits{Y.}},
\bauthor{\bsnm{Zhao}, \binits{S.}},
\bauthor{\bsnm{Feng}, \binits{X.-Q.}}:
\batitle{Chiral topographic instability in shrinking spheres}.
\bjtitle{Nat. Comput. Sci.}
\bvolume{2}(\bissue{10}),
\bfpage{632}--\blpage{640}
(\byear{2022})
\end{barticle}
\endbibitem

\bibitem[\protect\citeauthoryear{Striedter et~al.}{2015}]{striedter2015review}
\begin{barticle}
\bauthor{\bsnm{Striedter}, \binits{G.F.}},
\bauthor{\bsnm{Srinivasan}, \binits{S.}},
\bauthor{\bsnm{Monuki}, \binits{E.S.}}:
\batitle{Cortical folding: {W}hen, where, how, and why?}
\bjtitle{Annu. Rev. Neurosci.}
\bvolume{38},
\bfpage{291}--\blpage{307}
(\byear{2015})
\end{barticle}
\endbibitem

\bibitem[\protect\citeauthoryear{Budday and Steinmann}{2018}]{budday2018IJSS}
\begin{barticle}
\bauthor{\bsnm{Budday}, \binits{S.}},
\bauthor{\bsnm{Steinmann}, \binits{P.}}:
\batitle{On the influence of inhomogeneous stiffness and growth on mechanical
  instabilities in the developing brain}.
\bjtitle{Int. J. Solids Struct.}
\bvolume{132},
\bfpage{31}--\blpage{41}
(\byear{2018})
\end{barticle}
\endbibitem

\bibitem[\protect\citeauthoryear{da~Costa~Campos
  et~al.}{2021}]{da2021NeuroImage}
\begin{barticle}
\bauthor{\bsnm{Costa~Campos}, \binits{L.}},
\bauthor{\bsnm{Hornung}, \binits{R.}},
\bauthor{\bsnm{Gompper}, \binits{G.}},
\bauthor{\bsnm{Elgeti}, \binits{J.}},
\bauthor{\bsnm{Caspers}, \binits{S.}}:
\batitle{The role of thickness inhomogeneities in hierarchical cortical
  folding}.
\bjtitle{NeuroImage}
\bvolume{231},
\bfpage{117779}
(\byear{2021})
\end{barticle}
\endbibitem

\bibitem[\protect\citeauthoryear{Fischl and Dale}{2000}]{fischl2000PNAS}
\begin{barticle}
\bauthor{\bsnm{Fischl}, \binits{B.}},
\bauthor{\bsnm{Dale}, \binits{A.M.}}:
\batitle{Measuring the thickness of the human cerebral cortex from magnetic
  resonance images}.
\bjtitle{Proc. Natl. Acad. Sci. USA}
\bvolume{97}(\bissue{20}),
\bfpage{11050}--\blpage{11055}
(\byear{2000})
\end{barticle}
\endbibitem

\bibitem[\protect\citeauthoryear{Chang et~al.}{2007}]{chang2007abnormal}
\begin{barticle}
\bauthor{\bsnm{Chang}, \binits{B.S.}},
\bauthor{\bsnm{Duzcan}, \binits{F.}},
\bauthor{\bsnm{Kim}, \binits{S.}},
\bauthor{\bsnm{Cinbis}, \binits{M.}},
\bauthor{\bsnm{Aggarwal}, \binits{A.}},
\bauthor{\bsnm{Apse}, \binits{K.A.}},
\bauthor{\bsnm{Ozdel}, \binits{O.}},
\bauthor{\bsnm{Atmaca}, \binits{M.}},
\bauthor{\bsnm{Zencir}, \binits{S.}},
\bauthor{\bsnm{Bagci}, \binits{H.}}, \betal:
\batitle{The role of reln in lissencephaly and neuropsychiatric disease}.
\bjtitle{Am. J. Med. Genet. B Neuropsychiatr. Genet.}
\bvolume{144},
\bfpage{58}--\blpage{63}
(\byear{2007})
\end{barticle}
\endbibitem

\bibitem[\protect\citeauthoryear{Xu et~al.}{2010}]{xu2010JBE}
\begin{botherref}
\oauthor{\bsnm{Xu}, \binits{G.}},
\oauthor{\bsnm{Knutsen}, \binits{A.K.}},
\oauthor{\bsnm{Dikranian}, \binits{K.}},
\oauthor{\bsnm{Kroenke}, \binits{C.D.}},
\oauthor{\bsnm{Bayly}, \binits{P.V.}},
\oauthor{\bsnm{Taber}, \binits{L.A.}}:
Axons pull on the brain, but tension does not drive cortical folding.
J. Biomech. Eng.
\textbf{132}(7)
(2010)
\end{botherref}
\endbibitem

\bibitem[\protect\citeauthoryear{Dervaux and Amar}{2008}]{dervaux2008PRL}
\begin{barticle}
\bauthor{\bsnm{Dervaux}, \binits{J.}},
\bauthor{\bsnm{Amar}, \binits{M.B.}}:
\batitle{Morphogenesis of growing soft tissues}.
\bjtitle{Phys. Rev. Lett.}
\bvolume{101}(\bissue{6}),
\bfpage{068101}
(\byear{2008})
\end{barticle}
\endbibitem

\bibitem[\protect\citeauthoryear{Johnson}{1983}]{johnson1983constitutive}
\begin{botherref}
\oauthor{\bsnm{Johnson}, \binits{G.R.}}:
A constitutive model and data for materials subjected to large strains, high
  strain rates, and high temperatures.
Proc. 7th Inf. Sympo. Ballistics,
541--547
(1983)
\end{botherref}
\endbibitem

\bibitem[\protect\citeauthoryear{Seidt and Gilat}{2013}]{seidt2013plastic}
\begin{barticle}
\bauthor{\bsnm{Seidt}, \binits{J.}},
\bauthor{\bsnm{Gilat}, \binits{A.}}:
\batitle{Plastic deformation of 2024-{T}351 aluminum plate over a wide range of
  loading conditions}.
\bjtitle{Int. J. Solids Struct.}
\bvolume{50}(\bissue{10}),
\bfpage{1781}--\blpage{1790}
(\byear{2013})
\end{barticle}
\endbibitem

\bibitem[\protect\citeauthoryear{Ko and Fleischer}{2013}]{ko2013loadings}
\begin{botherref}
\oauthor{\bsnm{Ko}, \binits{W.L.}},
\oauthor{\bsnm{Fleischer}, \binits{V.T.}}:
Method for estimating operational loads on aerospace structures using
  span-wisely distributed surface strains.
Technical report
(2013)
\end{botherref}
\endbibitem

\bibitem[\protect\citeauthoryear{Clouchoux et~al.}{2012}]{clouchoux2012GI}
\begin{barticle}
\bauthor{\bsnm{Clouchoux}, \binits{C.}},
\bauthor{\bsnm{Kudelski}, \binits{D.}},
\bauthor{\bsnm{Gholipour}, \binits{A.}},
\bauthor{\bsnm{Warfield}, \binits{S.K.}},
\bauthor{\bsnm{Viseur}, \binits{S.}},
\bauthor{\bsnm{Bouyssi-Kobar}, \binits{M.}},
\bauthor{\bsnm{Mari}, \binits{J.-L.}},
\bauthor{\bsnm{Evans}, \binits{A.C.}},
\bauthor{\bsnm{Du~Plessis}, \binits{A.J.}},
\bauthor{\bsnm{Limperopoulos}, \binits{C.}}:
\batitle{Quantitative in vivo {MRI} measurement of cortical development in the
  fetus}.
\bjtitle{Brain Struct. Funct.}
\bvolume{217},
\bfpage{127}--\blpage{139}
(\byear{2012})
\end{barticle}
\endbibitem

\bibitem[\protect\citeauthoryear{Park and Friston}{2013}]{park2013Science}
\begin{barticle}
\bauthor{\bsnm{Park}, \binits{H.-J.}},
\bauthor{\bsnm{Friston}, \binits{K.}}:
\batitle{Structural and functional brain networks: from connections to
  cognition}.
\bjtitle{Science}
\bvolume{342}(\bissue{6158}),
\bfpage{1238411}
(\byear{2013})
\end{barticle}
\endbibitem

\bibitem[\protect\citeauthoryear{Zhao et~al.}{2022}]{zhao2022Patterns}
\begin{barticle}
\bauthor{\bsnm{Zhao}, \binits{Y.}},
\bauthor{\bsnm{Qin}, \binits{J.}},
\bauthor{\bsnm{Wang}, \binits{S.}},
\bauthor{\bsnm{Xu}, \binits{Z.}}:
\batitle{Unraveling the morphological complexity of two-dimensional
  macromolecules}.
\bjtitle{Patterns}
\bvolume{3},
\bfpage{100497}
(\byear{2022})
\end{barticle}
\endbibitem

\bibitem[\protect\citeauthoryear{Boucher et~al.}{2009}]{boucher2009depth}
\begin{barticle}
\bauthor{\bsnm{Boucher}, \binits{M.}},
\bauthor{\bsnm{Whitesides}, \binits{S.}},
\bauthor{\bsnm{Evans}, \binits{A.}}:
\batitle{Depth potential function for folding pattern representation,
  registration and analysis}.
\bjtitle{Med. Image Anal.}
\bvolume{13}(\bissue{2}),
\bfpage{203}--\blpage{214}
(\byear{2009})
\end{barticle}
\endbibitem

\bibitem[\protect\citeauthoryear{Fan et~al.}{2017}]{fan2017ChamferDisance}
\begin{bchapter}
\bauthor{\bsnm{Fan}, \binits{H.}},
\bauthor{\bsnm{Su}, \binits{H.}},
\bauthor{\bsnm{Guibas}, \binits{L.J.}}:
\bctitle{A point set generation network for 3{D} object reconstruction from a
  single image}.
In: \bbtitle{Proc. IEEE Comput. Soc. Conf. Comput. Vis. Pattern Recognit.},
pp. \bfpage{605}--\blpage{613}
(\byear{2017})
\end{bchapter}
\endbibitem

\bibitem[\protect\citeauthoryear{Grziwotz et~al.}{2023}]{grziwotz2023DEV}
\begin{barticle}
\bauthor{\bsnm{Grziwotz}, \binits{F.}},
\bauthor{\bsnm{Chang}, \binits{C.-W.}},
\bauthor{\bsnm{Dakos}, \binits{V.}},
\bauthor{\bsnm{Nes}, \binits{E.H.}},
\bauthor{\bsnm{Schwarzl{\"a}nder}, \binits{M.}},
\bauthor{\bsnm{Kamps}, \binits{O.}},
\bauthor{\bsnm{He{\ss}ler}, \binits{M.}},
\bauthor{\bsnm{Tokuda}, \binits{I.T.}},
\bauthor{\bsnm{Telschow}, \binits{A.}},
\bauthor{\bsnm{Hsieh}, \binits{C.-H.}}:
\batitle{Anticipating the occurrence and type of critical transitions}.
\bjtitle{Sci. Adv.}
\bvolume{9}(\bissue{1}),
\bfpage{4558}
(\byear{2023})
\end{barticle}
\endbibitem

\bibitem[\protect\citeauthoryear{Wang et~al.}{2020}]{wang2020Matter}
\begin{barticle}
\bauthor{\bsnm{Wang}, \binits{Y.}},
\bauthor{\bsnm{Wang}, \binits{S.}},
\bauthor{\bsnm{Li}, \binits{P.}},
\bauthor{\bsnm{Rajendran}, \binits{S.}},
\bauthor{\bsnm{Xu}, \binits{Z.}},
\bauthor{\bsnm{Liu}, \binits{S.}},
\bauthor{\bsnm{Guo}, \binits{F.}},
\bauthor{\bsnm{He}, \binits{Y.}},
\bauthor{\bsnm{Li}, \binits{Z.}},
\bauthor{\bsnm{Xu}, \binits{Z.}}, \betal:
\batitle{Conformational phase map of two-dimensional macromolecular graphene
  oxide in solution}.
\bjtitle{Matter}
\bvolume{3}(\bissue{1}),
\bfpage{230}--\blpage{245}
(\byear{2020})
\end{barticle}
\endbibitem

\bibitem[\protect\citeauthoryear{Keller}{1987}]{keller1987path-following}
\begin{barticle}
\bauthor{\bsnm{Keller}, \binits{H.B.}}:
\batitle{Lectures on numerical methods in bifurcation problems}.
\bjtitle{Appl. Math.}
\bvolume{217},
\bfpage{50}
(\byear{1987})
\end{barticle}
\endbibitem

\bibitem[\protect\citeauthoryear{Bury et~al.}{2021}]{bury2021PNAS}
\begin{barticle}
\bauthor{\bsnm{Bury}, \binits{T.M.}},
\bauthor{\bsnm{Sujith}, \binits{R.}},
\bauthor{\bsnm{Pavithran}, \binits{I.}},
\bauthor{\bsnm{Scheffer}, \binits{M.}},
\bauthor{\bsnm{Lenton}, \binits{T.M.}},
\bauthor{\bsnm{Anand}, \binits{M.}},
\bauthor{\bsnm{Bauch}, \binits{C.T.}}:
\batitle{Deep learning for early warning signals of tipping points}.
\bjtitle{Proc. Natl. Acad. Sci. USA}
\bvolume{118}(\bissue{39}),
\bfpage{2106140118}
(\byear{2021})
\end{barticle}
\endbibitem

\bibitem[\protect\citeauthoryear{Ni et~al.}{2023}]{ni2023Chem}
\begin{barticle}
\bauthor{\bsnm{Ni}, \binits{B.}},
\bauthor{\bsnm{Kaplan}, \binits{D.L.}},
\bauthor{\bsnm{Buehler}, \binits{M.J.}}:
\batitle{Generative design of de novo proteins based on secondary-structure
  constraints using an attention-based diffusion model}.
\bjtitle{Chem}
\bvolume{9},
\bfpage{1828}--\blpage{1849}
(\byear{2023})
\end{barticle}
\endbibitem

\bibitem[\protect\citeauthoryear{Luu et~al.}{2023}]{luu2023APL}
\begin{barticle}
\bauthor{\bsnm{Luu}, \binits{R.K.}},
\bauthor{\bsnm{Wysokowski}, \binits{M.}},
\bauthor{\bsnm{Buehler}, \binits{M.J.}}:
\batitle{Generative discovery of de novo chemical designs using diffusion
  modeling and transformer deep neural networks with application to deep
  eutectic solvents}.
\bjtitle{Appl. Phys. Lett.}
\bvolume{122},
\bfpage{234103}
(\byear{2023})
\end{barticle}
\endbibitem

\bibitem[\protect\citeauthoryear{Barak et~al.}{2011}]{barak2011NatureGenetics}
\begin{barticle}
\bauthor{\bsnm{Barak}, \binits{T.}},
\bauthor{\bsnm{Kwan}, \binits{K.Y.}},
\bauthor{\bsnm{Louvi}, \binits{A.}},
\bauthor{\bsnm{Demirbilek}, \binits{V.}},
\bauthor{\bsnm{Sayg{\i}}, \binits{S.}},
\bauthor{\bsnm{T{\"u}ys{\"u}z}, \binits{B.}},
\bauthor{\bsnm{Choi}, \binits{M.}},
\bauthor{\bsnm{Boyac{\i}}, \binits{H.}},
\bauthor{\bsnm{Doerschner}, \binits{K.}},
\bauthor{\bsnm{Zhu}, \binits{Y.}}, \betal:
\batitle{Recessive {LAMC3} mutations cause malformations of occipital cortical
  development}.
\bjtitle{Nat. Genet.}
\bvolume{43}(\bissue{6}),
\bfpage{590}--\blpage{594}
(\byear{2011})
\end{barticle}
\endbibitem

\bibitem[\protect\citeauthoryear{Poirier
  et~al.}{2013}]{poirier2013NatureGenetics}
\begin{barticle}
\bauthor{\bsnm{Poirier}, \binits{K.}},
\bauthor{\bsnm{Lebrun}, \binits{N.}},
\bauthor{\bsnm{Broix}, \binits{L.}},
\bauthor{\bsnm{Tian}, \binits{G.}},
\bauthor{\bsnm{Saillour}, \binits{Y.}},
\bauthor{\bsnm{Boscheron}, \binits{C.}},
\bauthor{\bsnm{Parrini}, \binits{E.}},
\bauthor{\bsnm{Valence}, \binits{S.}},
\bauthor{\bsnm{Pierre}, \binits{B.S.}},
\bauthor{\bsnm{Oger}, \binits{M.}}, \betal:
\batitle{Mutations in {TUBG1}, {DYNC1H1}, {KIF5C} and {KIF2A} cause
  malformations of cortical development and microcephaly}.
\bjtitle{Nat. Genet.}
\bvolume{45}(\bissue{6}),
\bfpage{639}--\blpage{647}
(\byear{2013})
\end{barticle}
\endbibitem

\bibitem[\protect\citeauthoryear{Morio and Balesdent}{2015}]{morio2015rare}
\begin{bbook}
\bauthor{\bsnm{Morio}, \binits{J.}},
\bauthor{\bsnm{Balesdent}, \binits{M.}}:
\bbtitle{Estimation of Rare Event Probabilities in Complex Aerospace and Other
  Systems: {A} Practical Approach}.
\bpublisher{Woodhead Publishing},
\blocation{Amsterdam}
(\byear{2015})
\end{bbook}
\endbibitem

\bibitem[\protect\citeauthoryear{Pickering
  et~al.}{2022}]{pickering2022rare_event}
\begin{barticle}
\bauthor{\bsnm{Pickering}, \binits{E.}},
\bauthor{\bsnm{Guth}, \binits{S.}},
\bauthor{\bsnm{Karniadakis}, \binits{G.E.}},
\bauthor{\bsnm{Sapsis}, \binits{T.P.}}:
\batitle{Discovering and forecasting extreme events via active learning in
  neural operators}.
\bjtitle{Nat. Comput. Sci.}
\bvolume{2}(\bissue{12}),
\bfpage{823}--\blpage{833}
(\byear{2022})
\end{barticle}
\endbibitem

\bibitem[\protect\citeauthoryear{Bi et~al.}{2023}]{bi2023pangu}
\begin{barticle}
\bauthor{\bsnm{Bi}, \binits{K.}},
\bauthor{\bsnm{Xie}, \binits{L.}},
\bauthor{\bsnm{Zhang}, \binits{H.}},
\bauthor{\bsnm{Chen}, \binits{X.}},
\bauthor{\bsnm{Gu}, \binits{X.}},
\bauthor{\bsnm{Tian}, \binits{Q.}}:
\batitle{Accurate medium-range global weather forecasting with 3{D} neural
  networks}.
\bjtitle{Nature}
\bvolume{619},
\bfpage{533}--\blpage{538}
(\byear{2023})
\end{barticle}
\endbibitem

\bibitem[\protect\citeauthoryear{Ericson}{2004}]{ericson2004contact}
\begin{bbook}
\bauthor{\bsnm{Ericson}, \binits{C.}}:
\bbtitle{Real-time Collision Detection}.
\bpublisher{CRC Press},
\blocation{Florida}
(\byear{2004})
\end{bbook}
\endbibitem

\bibitem[\protect\citeauthoryear{Belytschko et~al.}{2014}]{belytschko2014FEA}
\begin{bbook}
\bauthor{\bsnm{Belytschko}, \binits{T.}},
\bauthor{\bsnm{Liu}, \binits{W.K.}},
\bauthor{\bsnm{Moran}, \binits{B.}},
\bauthor{\bsnm{Elkhodary}, \binits{K.}}:
\bbtitle{Nonlinear Finite Elements for Continua and Structures}.
\bpublisher{John Wiley \& Sons},
\blocation{New Jersey}
(\byear{2014})
\end{bbook}
\endbibitem

\bibitem[\protect\citeauthoryear{Fan et~al.}{2017}]{fan2017MRI}
\begin{bchapter}
\bauthor{\bsnm{Fan}, \binits{B.}},
\bauthor{\bsnm{Rao}, \binits{Y.}},
\bauthor{\bsnm{Liu}, \binits{W.}},
\bauthor{\bsnm{Wang}, \binits{Q.}},
\bauthor{\bsnm{Wen}, \binits{H.}}:
\bctitle{Region-based growing algorithm for 3{D} reconstruction from {MRI}
  images}.
In: \bbtitle{Int. Conf. Image Vis. Comput.},
pp. \bfpage{521}--\blpage{525}
(\byear{2017})
\end{bchapter}
\endbibitem

\bibitem[\protect\citeauthoryear{Qi et~al.}{2017}]{qi2017pointnet}
\begin{bchapter}
\bauthor{\bsnm{Qi}, \binits{C.R.}},
\bauthor{\bsnm{Su}, \binits{H.}},
\bauthor{\bsnm{Mo}, \binits{K.}},
\bauthor{\bsnm{Guibas}, \binits{L.J.}}:
\bctitle{Pointnet: {D}eep learning on point sets for 3{D} classification and
  segmentation}.
In: \bbtitle{Proc. IEEE Comput. Soc. Conf. Comput. Vis. Pattern Recognit.},
pp. \bfpage{652}--\blpage{660}
(\byear{2017})
\end{bchapter}
\endbibitem

\bibitem[\protect\citeauthoryear{Achlioptas et~al.}{2018}]{achlioptas2018ICML}
\begin{bchapter}
\bauthor{\bsnm{Achlioptas}, \binits{P.}},
\bauthor{\bsnm{Diamanti}, \binits{O.}},
\bauthor{\bsnm{Mitliagkas}, \binits{I.}},
\bauthor{\bsnm{Guibas}, \binits{L.}}:
\bctitle{Learning representations and generative models for 3{D} point clouds}.
In: \bbtitle{Int. Conf. Mach. Learn.},
pp. \bfpage{40}--\blpage{49}
(\byear{2018})
\end{bchapter}
\endbibitem

\bibitem[\protect\citeauthoryear{Ushio et~al.}{2018}]{ushio2018Nature}
\begin{barticle}
\bauthor{\bsnm{Ushio}, \binits{M.}},
\bauthor{\bsnm{Hsieh}, \binits{C.-h.}},
\bauthor{\bsnm{Masuda}, \binits{R.}},
\bauthor{\bsnm{Deyle}, \binits{E.R.}},
\bauthor{\bsnm{Ye}, \binits{H.}},
\bauthor{\bsnm{Chang}, \binits{C.-W.}},
\bauthor{\bsnm{Sugihara}, \binits{G.}},
\bauthor{\bsnm{Kondoh}, \binits{M.}}:
\batitle{Fluctuating interaction network and time-varying stability of a
  natural fish community}.
\bjtitle{Nature}
\bvolume{554}(\bissue{7692}),
\bfpage{360}--\blpage{363}
(\byear{2018})
\end{barticle}
\endbibitem

\bibitem[\protect\citeauthoryear{Cenci and Saavedra}{2019}]{cenci2019EDM}
\begin{barticle}
\bauthor{\bsnm{Cenci}, \binits{S.}},
\bauthor{\bsnm{Saavedra}, \binits{S.}}:
\batitle{Non-parametric estimation of the structural stability of
  non-equilibrium community dynamics}.
\bjtitle{Nat. Ecol. Evol.}
\bvolume{3}(\bissue{6}),
\bfpage{912}--\blpage{918}
(\byear{2019})
\end{barticle}
\endbibitem

\bibitem[\protect\citeauthoryear{Kuznetsov
  et~al.}{1998}]{kuznetsov1998bifurcation}
\begin{bbook}
\bauthor{\bsnm{Kuznetsov}, \binits{Y.A.}},
\bauthor{\bsnm{Kuznetsov}, \binits{I.A.}},
\bauthor{\bsnm{Kuznetsov}, \binits{Y.}}:
\bbtitle{Elements of Applied Bifurcation Theory}
vol. \bseriesno{112}.
\bpublisher{Springer},
\blocation{New {Y}ork}
(\byear{1998})
\end{bbook}
\endbibitem

\end{thebibliography}
\clearpage
\newpage 

\section*{Acknowledgments}

This study was supported by the National Natural Science Foundation of China through grants 11825203, 11832010, 11921002, and 52090032.
The computation was performed on the Explorer 100 cluster system of the Tsinghua National Laboratory for Information Science and Technology.

\section*{Author contributions}

Z.X. conceived and supervised the research.
Y.Z. performed the simulations and analysis.
Both authors wrote the manuscript.

\section*{Competing interests}

The authors declare that they have no competing financial interests. 

\section*{Additional information}
\noindent{\bf Supplementary information}
The online version contains supplementary material available.
\\
\noindent{\bf Correspondence and requests for materials}
should be addressed to Z.X. (xuzp@tsinghua.edu.cn).
\end{document}